\documentclass [12pt]{article}
\usepackage{amsmath,amssymb}

\usepackage{curves,epic,eepic,graphics}

\usepackage{graphics}
\usepackage{epsfig}
\usepackage{graphicx}
\usepackage{pgf}
\usepackage{pgfpages}
\usepackage{xcolor}

\usepackage{color}

\usepackage{pstricks}
\usepackage{tikz}
\linethickness{9pt}   
\usepackage{tkz-euclide}
\usetkzobj{all}
\usetikzlibrary{calc}

\begin{document}
\title{\textbf{\large Symmetry of Generalized Randall-Sundrum Model and Distribution of 3-Branes in Six-Dimensional Spacetime}}
\author{Sheng-Fei FENG$^{1,2}$ \hspace{0.1cm} Chang-Yu HUANG$^{3}$ \hspace{0.1cm} Yong-Chang HUANG$^{1}$ \\
 Xin LIU$^{1,4}\thanks{Corresponding author: xin.liu@bjut.edu.cn}$ \hspace{0.2cm}Ying-Jie ZHAO$^{1}$ \\
\\ {\small 1. Institute of Theoretical Physics, Beijing University of Technology,  Beijing, 100124, China}\\
 {\small 2.Department of Physics, Capital Normal University, Beijing, 100048, China}\\
 {\small 3. Department of Physics, Purdue University, 525 Northwestern Avenue, W. Lafayette, IN 47907-2036, USA}\\
 {\small 4. Beijing-Dublin International College, Beijing University of Technology,  Beijing, 100124, China}\\
 }
\date{}
\maketitle

\abstract{A generalization from the usual $5$-dimensional two-brane Randall-Sundrum (RS) model to a $6$-dimensional multi-brane RS model is presented. The extra dimensions are extended from one to two; correspondingly the single-variable warp function is generalized to be a double-variable function, to represent the two extra dimensions. In the analysis of the Einstein equation we have two remarkable discoveries. One is that, when branes are absent, the cosmological parameter distributed in the two extra dimensions acts as a function describing a family of circles. These circles are not artificially added ones but stem from the equations of motion, while their radii are inversely proportional to the square root of the cosmological parameter. The other discovery is that, on any circle, there symmetrically distribute four branes. Their tensions, $V_1 \sim V_4$, satisfy a particular relationship
$V_1=V_3=-V_2=-V_4=3M^4$, where $M$ is the $6$-dimensional fundamental scale of the RS model.}

\parsep=1mm \parskip=1mm

\begin{section}
{Introduction}
\end{section}
In 1999 Randall and Sundrum (RS) proposed a $5$-dimensional two brane model for solving the gauge hierarchy problem between the Planck and the electro-weak scales \cite{RSmodels}. In the RS scenario the background spacetime is an $AdS_5 $ bulk, in which there exist two $3$-dimensional branes with opposite tensions. That RS model has a symmetry of  $S^1 / \mathbb{Z}_2 $, and the exponential warp factor in the spacetime metric is able to generate a scale hierarchy on the ``$-$'' brane (the brane with negative tension) --- that is, if a standard model (SM) field is located on the ``$-$'' brane, it leads to a resolution of the gauge hierarchy problem. However, this theory, although sounds plausible in string theory, contradicts the usual knowledge that our visible world should reside on a ``$+$'' tension brane. In order to circumvent this difficulty many efforts of modification have been made. In Ref.\cite{LykkenRandall2000} Lykken and Randall suggested a ``$++-$ '' multi-brane model, in which the SM fields reside on the intermediate ``$+$'' brane, with a warp factor responsible for the scale hierarchy. Kogan \textit{et al.} in Ref.\cite{Koganetal2001} considered a ``$+ - - +$'' multi-brane model and a so-called crystal universe model  \cite{Koganetal2000,Gregoryetal2000,Oda2000,Hatanakaetal1999,Li2000PLB,Kaloper2000,Li2000NPB}. These provide a way to put the visible sector on a ``+'' brane with
an hierarchical warp factor, demonstrating an interesting possibility that gravity could be different from expected not only at a small scale but also at an ultralarge scale. As pointed out by Arkani-Hames \textit{et al.} \cite{Arkani-Hamesetal2000}, the above models are able to give rise to many new experimentally testable phenomena such as neutrino mixing, dark matter and so on. Recent efforts in this area can be found in Refs.\cite{BalcerzakDabrowski2011,Alexeyevetal2015,NevesMolina2012,DasSenGupta2011,ReeceWang2010,
Liuetal2014,Jonesetal2013,BambaNojiriOdintsov2013}.

Another weak point of the $5$-dimensional RS scenario is instability. In order to produce stability Goldberger and Wise introduced a bulk scalar field, which minimizes the
potential generated by the bulk scalar and keeps all quartic interactions localized within two 3-branes \cite{GoldbergerWise1999}. Moreover, the Goldberger-Wise mechanism is generalized to a one-dimensional crystal universe model in Ref.\cite{Piaoetal2001}, leading to a consequence that the brane crystal is not equidistant in the absence of fine-tuning, and any ``$- +$'' pair of branes is far away from the adjacent ``$- +$'' pair in general.

In order to overcome the above difficulties, we propose in this paper a generalized $6$-dimensional multi-brane RS model with $2$ extra dimensions. In Sect.2, the action of the proposed model and the Einstein equation will be presented. In Sects.3 and 4, the solutions of the Einstein equation --- without and with branes, respectively --- will be discussed.


\begin{section}
{Action and Einstein equation}
\end{section}

We begin with presenting the origin of our RS model in the $6$-dimensional spacetime. With symmetry in 6D, a warp function should depend on two variables; correspondingly, our proposed 3-dimensional branes are distributed in these two extra dimensions.

The generic action for this configuration is given by
\begin{equation}
S = \int {d^4} x\int {dydz\sqrt { - \tilde {g}} } (2M^4\tilde {R}
- \tilde {\Lambda }) - \sum\limits_{i}\int_{\begin{subarray}{1}y =
y_i\\
z = z_i
\end{subarray}}
{d^4xV_{i} \sqrt { - g^{(i)}} }, \label{GenericAction}
\end{equation}

\noindent where $g_{\mu \nu }^{(i)} $ denotes the induced metric on the branes, $\mu,\nu = 1,2,3,4$. The $V_{i} $ denotes the tensions of the branes, and $M$ the 6D fundamental scale. $\tilde {\Lambda }$ is a six-dimensional cosmological constant parameter (--- in the next section, however, $\tilde {\Lambda }$ will no longer be a constant, but a variable). The integration of the last term of (\ref{GenericAction}) is conducted within the branes located on the $yz$ plane. The doublet $(y_i ,z_i )$, $i =1,2, \cdots, n$, denotes the coordinates of the branes in the two extra dimensions. For convenience let us denote the two extra dimensions as $y$ and $z$; in this paper a target we wish to achieve is to solve exact evaluation for $(y_i,z_i)$.

In the $N$-dimensional spacetime a general line element satisfying the 4-dimensional Poincare invariance reads \cite{GherghettaShaposhnikov2000}
\begin{equation}
ds^2=\gamma (x^a)g_{\mu \nu }(x^\mu )dx^\mu dx^\nu  +h_{ab}(x^a)dx^adx^b,\ \ \ \ \ \ a,b=5,\cdots,N,
\end{equation}
where $h_{ab}(x^a)$ is the metric associated with the $(N-4)$ extra dimensions, and
$\gamma (x^a)$ a conformal factor depending uniquely on the extra coordinates.
In contrast to the usual $5$-dimensional RS model with only one extra dimension, what we consider here is a $6$-dimensional spacetime with a symmetry assumed between the two extra dimensions, such that $h_{ab}(x^a)$ are kept diagonal: $h_{11}(x^a)=h_{22}(x^a)=h(x^a)$. Then the line element is given by
\begin{equation}
ds^2=\gamma (y,z)g_{\mu \nu }(x^\mu )dx^\mu dx^\nu +h(y,z)(dy^2+dz^2).
\end{equation}

In fact, there have been so far much work focusing on codimension-two brane-worlds \cite{GherghettaShaposhnikov2000,Burgessetal2002,Multamaki2002,Chodos1999,
CohenKaplan1999,Gregory2000,OlasagastiVilenkin2000,Csakietal2000}. In this paper our aim is to investigate the origin of the symmetry for a 6D RS model. For this purpose we start from a conformal transformation
\begin{equation}
G'_{AB}=
h^{-1}(y,z)G_{AB},\ \ \ \ \ \ A,B=\{\mu ,a\},\ \ \ \ \text{i.e. }\ \ \ \
h'_{ab}=h^{-1}(y,z)h_{ab}, G'_{\mu\nu}= h^{-1}(y,z)\gamma
(y,z)g_{\mu\nu}.
\end{equation}
Then the line element is rewritten as
\begin{equation}
ds^2 = e^{ - 2\sigma (y,z)}g _{\mu \nu } dx^\mu dx^\nu + dy^2 +
dz^2, \label{Metric0}
\end{equation}
where the signature of the metric is $(-,+,+,+,+,+)$, and $e^{ - 2\sigma (y,z)}=h^{-1}(y,z)\gamma (y,z)$. The warp function $\sigma (y,z)$ is essentially a conformal factor rescaling the $4$-dimensional components of the metric. The Einstein equation now reads
\begin{eqnarray}
 &&\sqrt{ - \tilde {g}} 2M^4(\tilde{R}_{mn} - \frac{1}{2}\tilde {g}_{mn}
\tilde {R}) + \frac{1}{2}\tilde {\Lambda }\tilde {g}_{mn} \sqrt{- \tilde {g}} \nonumber\\
&&\hspace{2cm} + \frac{1}{2}\sum\limits_{i} {V_{i} \sqrt { -
g_{(i)} } g_{(i)\mu \nu } \delta _m^{\mu} \delta _n^{\nu} \delta (y -
y_i )\delta (z - z_i )} = 0,    \label{EinsteinEqn}
 \end{eqnarray}
where $m$ and $n$ range from $1$ to $6$, with ``$5$''
corresponding to the fifth coordinate $y$ and ``$6$'' to the sixth $z$.
Under the metric (\ref{Metric0}) the non-vanishing components of the connection are
 \begin{eqnarray}
\Gamma _{\mu5}^\mu &=& - \sigma _{,y} (y,z),  \label{Connxn-1}\\
\Gamma _{\mu6}^\mu &=& -\sigma _{,y} (y,z),  \label{Connxn-2}\\
\Gamma _{\mu\mu}^5 &=& e^{ - 2\sigma (y,z)}\sigma
_{,y} (y,z),  \label{Connxn-3}\\
\Gamma _{\mu\mu}^6 &=& - e^{ - 2\sigma (y,z)}\sigma
_{,z} (y,z),  \label{Connxn-4}
\end{eqnarray}
where no summation convention applies upon repeated indices. Correspondingly, the non-vanishing components of the curvature are
\begin{eqnarray}
-\tilde {R}_{11} &=& \tilde {R}_{22} = \tilde {R}_{33} = \tilde
 {R}_{44}   \nonumber\\
 &=& - 4e^{ - 2\sigma (y,z)}\sigma _{,y} ^2(y,z) + e^{ - 2\sigma (y,z)}\sigma
_{,y} (y,z)   \nonumber\\
 && -4e^{ - 2\sigma (y,z)}\sigma _{,z} ^2(y,z) + e^{ - 2\sigma
(y,z)}\sigma _{,z} (y,z), \label{Curvature-1}\\
\tilde {R}_{55} &=& 4\sigma _{,yy} (y,z) - 4\sigma _{,y} ^2(y,z), \label{Curvature-2}\\
\tilde {R}_{66} &=& 4\sigma _{,z} (y,z) - 4\sigma _{,z}
^2(y,z), \label{Curvature-3}\\
\tilde {R}_{56} &=& \sigma _{,yz} (y,z) - \sigma
_{,y} (y,z)\sigma _{,z} (y,z). \label{Curvature-4}
\end{eqnarray}
Substituting the connection and curvature into the Einstein equation (\ref{EinsteinEqn}) we have
\begin{eqnarray}
&&{\sigma }_{,yz} (y,z) - {\sigma }_{,y} (y,z){\sigma }_{,z} (y,z) =
 0,     \label{ConformalFact-1}\\
&&- 6{\sigma }^2_{,y}(y,z) + 3{\sigma }_{,yy} (y,z) - 6{\sigma
}^2_{,z}(y,z) + 3{\sigma }_{,zz} (y,z)\nonumber \\
&&\hspace{2cm}= \frac{1}{4M^4}[\tilde {\Lambda } + \sum\limits_{i}
{V_{i}\delta (y - y_i )\delta (z - z_i )} ],     \label{ConformalFact-2}\\
&&6{\sigma }^2_{,y} (y,z) +
10{\sigma }^2_{,z} (y,z) - 4{\sigma }_{,zz} (y,z) =
-\frac{1}{4M^4}\tilde {\Lambda },     \label{ConformalFact-3}\\
&&6{\sigma }^2_{,z} (y,z) + 10{\sigma }^2_{,y} (y,z) - 4{\sigma
}_{,yy} (y,z) = - \frac{1}{4M^4}\tilde {\Lambda }.     \label{ConformalFact-4}
\end{eqnarray}
Comparing (\ref{ConformalFact-1})--(\ref{ConformalFact-4}) with the counterpart of the one-dimensional crystal universe model \cite{Piaoetal2001}, Eq.(\ref{ConformalFact-1}) is recognized to be a new condition. We need to find a solution to (\ref{ConformalFact-1}) which is compatible with (\ref{ConformalFact-2})--(\ref{ConformalFact-4}) simultaneously.

A solution we have found is
\begin{equation}
\sigma (y,z) = - \ln \left(f(y)\pm g\left(z\right) + c_1 \right) + c_2.    \label{sigmayz-pm}
\end{equation}
The ``$-$'' branch of (\ref{sigmayz-pm}) should be abandoned, because it speaks of anti-symmetry but the coordinates $y$ and $z$ are actually symmetric. Thus
\begin{equation}
\sigma (y,z) = - \ln \left(f(y) + g\left(z\right) + c_1 \right) + c_2. \label{ConformalFact-condn}
\end{equation}
Substituting (\ref{ConformalFact-condn}) into (\ref{ConformalFact-2})--(\ref{ConformalFact-4}) we have
\begin{eqnarray}
   &&{\frac{3{f}'^2(y) - 3{f}''(y)(f(y) + g(z) + c_1
) + 3{g}'^2(z) - 3{g}''(z)(f(y) + g(z) + c_1 )}{(f(y) + g(z) + c_1
)^2}}\nonumber\\
 &&\hspace{2cm}{+\frac{ - 6{f}'^2(y) - 6{g}'^2(z)}{(f(y) + g(z) + c_1 )^2}
 = \frac{1}{4M^4}[\tilde {\Lambda } + \sum\limits_{i} {V_{i} \delta (y -
y_i )\delta (z - z_i)} ]},     \label{condn-1}\\
 &&{\frac{6{f}'^2(y) + 6{g}'^2(z) +
4{f}''(y)(f(y) + g(z) + c_1 )}{(f(y) + g(z) + c_1 )^2} = -
\frac{1}{4M^4}\tilde {\Lambda } },     \label{condn-2}\\
 &&{\frac{6{f}'^2(y) +
6{g}'^2(z) + 4{g}''(z)(f(y) + g(z) + c_1 )}{(f(y) + g(z) + c_1
)^2} = - \frac{1}{4M^4}\tilde {\Lambda }}.     \label{condn-3}
\end{eqnarray}
This set of equations (\ref{condn-1})--(\ref{condn-3}) forms the base of our theory; further discussion will be on the top of it in the coming sections.


\begin{section}
{Solutions without branes}
\end{section}

Before studying the brane-world, let us first examine the possible static space-time configurations determined by the Einstein equation. Without the brane term, Eq.(18) is simplified as
\begin{eqnarray}
&&{\frac{3{f}'^2(y) - 3{f}''(y)(f(y) + g(z) + c_1
) + 3{g}'^2(z) - 3{g}''(z)(f(y) + g(z) + c_1 )}{(f(y) + g(z) + c_1
)^2}}\nonumber\\
&&\hspace{2cm}{+\frac{ - 6{f}'^2(y) - 6{g}'^2(z)}{(f(y) + g(z) + c_1 )^2}
 = \frac{1}{4M^4}\tilde {\Lambda } }.   \label{EinEqnWithoutbraneterm}
\end{eqnarray}
In terms of (\ref{condn-2}) and (\ref{condn-3}) we obtain $f''(y) =
g''(z) \neq 0$, which implies the existence of direct relevance between $y$ and $z$. Due to the symmetry between $y$ and $z$, we have $f''(y) = g''(z) = 2c$, then
\begin{equation}
f(y) = cy^2 + ay, g(z) = cz^2 + bz.   \label{fygy}
\end{equation}
Substituting (\ref{fygy}) into (\ref{condn-2})--(\ref{EinEqnWithoutbraneterm}) there is
\begin{equation}
(y + \frac{a}{2c})^2 + (z + \frac{b}{2c})^2 + \frac{a^2 + b^2 -
4cc_1 }{8c^2} = 0.    \label{CircleInduced}
\end{equation}
Eqs.(\ref{condn-2})--(\ref{EinEqnWithoutbraneterm}) are compatible equations because of (\ref{CircleInduced}). It is seen that (\ref{CircleInduced}) gives a circle centered at $\left(-\frac{a}{2c},-\frac{b}{2c}\right)$, that means, only the points on the circle are able to share the same cosmological parameter $\tilde {\Lambda }$. Substituting (\ref{CircleInduced}) into (\ref{EinEqnWithoutbraneterm}), we steadily obtain the radius of the circle
\begin{equation}
r = \sqrt {\frac{4cc_1 -a^2-b^2  }{8c^2}}= 8\sqrt { -
\frac{M^4}{3\tilde {\Lambda }}}.    \label{radiusLambda}
\end{equation}
Eq.(\ref{radiusLambda}) has clear physical meaning: in the plane of the two extra dimensions, the cosmological parameter is distributed on a family of circles, the radii of which are inversely proportional to the square root of the cosmological parameter. Moreover, in the usual $5$-dimensional RS model, the $S^1 / \mathbb{Z}_2 $ orbifold geometry of the extra dimension is added by hand, but in our generalized $6$-dimensional RS model this symmetry is strictly derived from the Einstein equation.

Furthermore, if we are about to find the branes sharing the same cosmological parameters in the six dimensional spacetime, we have to focus on the circles defined by (\ref{CircleInduced}). See below.


\begin{section}
{Branes}
\end{section}

We are now at the stage to study the brane world given by (\ref{condn-1})--(\ref{condn-3}). In terms of (\ref{fygy}) which stems from (\ref{condn-2}) and (\ref{condn-3}) we have
\begin{eqnarray}
&&\frac{ - 24c^2y^2 - 24c^2z^2 - 24acy - 24bcz - 3a^2 - 3b^2 - 12cc_1 }{(cy^2
+ ay + cz^2 + bz + c_1 )^2} \nonumber\\
&&\hspace{2cm}= \frac{1}{4M^4}[\tilde {\Lambda } + \sum\limits_{i} {V_{i} \delta (y -
y_i )\delta (z - z_i )} ], \label{withBrane-1}\\
&&\frac{32c^2y^2 + 32c^2z^2 + 32acy + 32bcz + 6a^2 + 6b^2 + 8cc_1 }{(cy^2 +
ay + cz^2 + bz + c_1 )^2} = - \frac{1}{4M^4}\tilde {\Lambda } .\label{withBrane-2}
\end{eqnarray}
For convenience let us denote $\xi=cy^2 + ay + cz^2 + bz
+ c_1 $ and $D=\frac{a^2 + b^2 - 4cc_1}{8c}$.
Then (\ref{withBrane-1}) and (\ref{withBrane-2}) become
\begin{eqnarray}
\frac{\xi + 3D / 2 - D / 2}{\xi ^2} &=& - \frac{1}{96cM^4}\tilde {\Lambda } - \frac{1}{96cM^4}\sum\limits_{i} {V_{i} \delta (y - y_i
)\delta (z - z_i)}, \\
\frac{\xi + 3D / 2}{\xi ^2} &=& - \frac{1}{128cM^4}\tilde {\Lambda },
\end{eqnarray}
which immediately lead to
\begin{eqnarray}
&&\frac{1}{\left[(y + \frac{a}{2c})^2 + (z + \frac{b}{2c})^2 - \frac{a^2 +
 b^2-4cc_1}{4c^2}\right] ^2}\nonumber\\
&&\hspace{2cm}=\frac{c}{192M^4D}\tilde {\Lambda
 }+\frac{c}{48M^4D}\sum\limits_{i} {V_{i} \delta (y - y_i
)\delta (z - z_i )}.   \label{Brane-orig29}
\end{eqnarray}

Seemingly Eq.(\ref{Brane-orig29}) has no solutions, but this is not true. At the first glance the singularities on the LHS of (\ref{Brane-orig29}) are poles of second order whereas those on the RHS are in a form of delta-functions, which denies the existence of solutions. However, we will show through calculations below that the LHS of (\ref{Brane-orig29}) can definitely be expressed in a form of double-delta-functions in agreement to the RHS.

On the other hand, it has been shown in our past work that the so-called quantitative causal principle should hold in general in physics \cite{Huangetal2006,Huangetal20042007,HuangWeng2005,
Huangetal20072007,Huangetal2002,Huangetal2013,ZhangHuang2011}. For instance, in \cite{Huangetal2002} the no-loss-no-gain homeomorphic map transformation is used to derive exact strain tensor formulae on a Weitzenb\"{o}ck manifold. In the light of this principle, we argue that a change of a quantity in the RHS of (\ref{Brane-orig29}) (which acts as a cause) must result in a change within another quantity in the LHS of (\ref{Brane-orig29}) (which serves as a consequence) --- this confirms the ``no-loss-no-gain'' rule. Since the RHS of (\ref{Brane-orig29}) has delta functions, there must exist delta functions in the LHS as well.

Let us introduce $r_{b} = \sqrt {\frac{a^2 + b^2 -
4cc_1 }{4c^2}}$, where $r_{b}$ means $r_{\text{brane}}$, and define $y'=y+
\frac{a}{2c}$ and $z'=z+\frac{b}{2c}$. Then in the two-dimensional $y'z'$-plane one can introduce the polar coordinates $\left(r,\theta\right)$ as
\begin{equation}
y'=r\cos\theta,\hspace*{15mm}z'=r\sin\theta.
\end{equation}
Eq.(\ref{Brane-orig29}) now becomes
\begin{equation}
\frac{1}{(r^2-r_{b}^2)^2}=  \frac{1}{96M^4}\frac{1}{r_{b}^2}\tilde {\Lambda }
 +\frac{1}{r_{b}^2}\frac{1}{24M^4}\sum\limits_{i} {V_{i}\delta (y'-\frac{a}{2c} - y_i )\delta (z'-\frac{b}{2c}- z_i )},     \label{Brane-orig30}
\end{equation}
where $D=cr^2_b/2$ applies. A notation $V_{i} \delta (y - y_i )\delta (z - z_i )$ is introduced to express the positions of the branes, due to a symmetry $r^2=\vec{r}\cdot\vec{r}=(-\vec{r})\cdot(-\vec{r})$.

The singularities of the both sides of (\ref{Brane-orig30}) should have one-to-one correspondence. This requires the branes to be symmetrically distributed on the circle --- that is, they must distribute in pair on the two opposite ends of a diameter. In detail, we have the following cases:
\begin{enumerate}

\item[Case 1:] When $r\neq r_{b}$, the LHS of (\ref{Brane-orig30}) has no singular points. Hence the RHS must have no singular points, meaning that $\frac{1}{24M^4}\sum\limits_{i} {V_{i}\delta (y'-\frac{a}{2c} - y_i )\delta (z'-\frac{b}{2c}- z_i )}=0$. Thus
\begin{equation}
\frac{1}{(r^2-r^2_{b})^2}=\frac{1}{96M^4}\frac{1}{r^2_{b}}\tilde {\Lambda }.   \label{Case1}
\end{equation}


\item[Case 2:] When $r=r_b$, there exist brane structures in (\ref{Brane-orig30}). There are two subcases:

\begin{enumerate}

\item[Subcase 1:]
    Singular points exist. A technical obstacle we encounter here is how to expand the LHS of (\ref{Brane-orig30}) to be a combination of a number of delta functions, such that the both sides of (\ref{Brane-orig30}) could be compared. To solve this difficulty we firstly rewrite the LHS of (\ref{Brane-orig30}) as a sum of separate terms
    \begin{equation}
       \frac{1}{(r^2-r^2_{b})^2}=\frac{1}{4r^2_{b}}\left[\frac{1}{(r-r_{b})^2}
       -\frac{2}{(r-r_{b})(r+r_{b})}+\frac{1}{(r+r_{b})^2}\right].  \label{Case2Subcase0}
    \end{equation}
    A typical method to tackle (\ref{Case2Subcase0}) is to do a replacement $y'=z'=\frac{r}{\sqrt{2}}$ and $r'_{b}=\frac{r_{b}}{\sqrt{2}}$ in the space of the extra dimensions, such that
    \begin{eqnarray}
        \frac{1}{(r^2-r^2_{b})^2}        &=&\frac{1}{16r'^2_{b}}\bigg[\frac{\delta_{y',r'_{b}}\delta_{z',r'_{b}}}{(y'-r'_{b})(z'-r'_{b})}
       -\frac{\delta_{y',r'_{b}}\delta_{z',-r'_{b}}}{(y'-r'_{b})(z'+r'_{b})} \nonumber \\
       &&\hspace*{15mm}  -\frac{\delta_{y',-r'_{b}}\delta_{z',r'_{b}}}{(y'+r'_{b})(z'-r'_{b})}
       +\frac{\delta_{y',-r'_{b}}\delta_{z',-r'_{b}}}{(y'+r'_{b})(z'+r'_{b})}\bigg]   \label{Case2Subcase1}
    \end{eqnarray}

    Now let us borrow a trick from complex analysis
    \begin{equation}
    \frac{1}{x} = \frac{1}{x+i\varepsilon} = \mathcal{P}\left(\frac{1}{x}\right)-i\pi\delta(x),\hspace*{10mm}
    x,\varepsilon \in \mathbb{R};\ \varepsilon \ll 1.
    \end{equation}
     This is a sort of complex analytical extension to deal with the singularity of $\frac1x$, where $\mathcal{P}$ indicates the principal value part and $i\pi\delta(x)$ the singular part. Then, any factor in (\ref{Case2Subcase1}) is able to be re-expressed as delta functions, say,
    \begin{equation}
    \frac{\delta_{y',r'_{b}}}{y'-r'_{b}} = \frac{1}{4M^2r_{b}}\sqrt{\frac{\tilde{\Lambda}}{6}} -i\pi \delta \left(y'-r'_{b}\right),
    \end{equation}
    where the first term acts as the principal part. Thus, the singular part of the LHS of (\ref{Case2Subcase1}) becomes
    \begin{eqnarray}
    &&\left.\frac{1}{(r^2-r^2_{b})^2} \right|_{\text{Singular part}}  \nonumber \\
    &=& -\frac{\pi ^2}{16r'^2_{b}}
    \bigg[\delta(y'-r'_{b})\delta(z'-r'_{b})-\delta(y'+r'_{b})\delta(z'-r'_{b})\nonumber\\
    &&\hspace{15mm}+\delta(y'+r'_{b})\delta(z'+r'_{b})
    -\delta(y'-r'_{b})\delta(z'+r'_{b})\bigg].
    \label{Brane-orig31}
    \end{eqnarray}
    It is noted that in this calculation any term which contains the product of a principal part and a singular delta function should vanish precisely because of the symmetry between $y',-y'$ and between $z',-z'$. This agrees to the fact that this type of terms are absent from the RHS of (\ref{Brane-orig30}).

    Then we can finally compare (\ref{Brane-orig31}) with the RHS of (\ref{Brane-orig30}). It is seen that the summation index can only take four value, $i=1,2,3,4$, hence
      \begin{equation}
      \frac{1}{16r'^2_{b}}=\frac{1}{r^2_{b}}\frac{1}{24M^4}V_1=\frac{1}{r^2_{b}}\frac{1}{24M^4}V_3
      =-\frac{1}{r^2_{b}}\frac{1}{24M^4}V_2=-\frac{1}{r^2_{b}}\frac{1}{24M^4}V_4,
      \end{equation}
      i.e.,
      \begin{equation}
      V_1=V_3=-V_2=-V_4=3M^4.
      \end{equation}
    This means the branes $V_1 \sim V_4$ are located in the $y'z'$-plane in an anti-clockwise order with $\pi/2$ in between, as shown in Figure \ref{Fig1}:
    \begin{eqnarray}
    V_1 : y'_1 =r'_b,z'_1 =r'_b;&\hspace*{15mm}& V_2 : y'_2= -r'_b,z'_2 =r'_b;\\
    V_3 : y'_3= -r'_b,z'_3 =-r'_b;&\hspace*{15mm}& V_4 : y'_4= r'_b,z'_4 =-r'_b.\\
    \end{eqnarray}
    Our real world may reside in $V_2$ or $V_4$. The brane pairs ($V_1,V_2$), ($V_2,V_3$), ($V_3,V_4$) and ($V_4,V_1$) are identical.
    \setlength{\unitlength}{3.5mm}
    \begin{figure}
    \begin{center}
       \begin{tikzpicture}[thick,scale=0.8]
          \draw[line width=1.0pt,->,>=latex] (-4,0) -- (4,0) node[anchor=south] {$y'$};
          \draw[line width=1.0pt,->,>=latex] (0,-4) -- (0,4) node[anchor=west] {$z'$};
          \draw (0,0) circle (2.5);
          \coordinate[label=45:$V_1$](V1)at(1.768,1.768);
          \coordinate[label=135:$V_2$](V2)at(-1.768,1.768);
          \coordinate[label=-135:$V_3$](V3)at(-1.768,-1.768);
          \coordinate[label=-45:$V_4$](V4)at(1.768,-1.768);

          \draw (V1) circle (0.15);
          \draw [fill] (V2) circle (0.15);
          \draw (V3) circle (0.15);
          \draw [fill](V4) circle (0.15);
          \draw (V1)--(V2);
          \draw (V2)--(V3);
          \draw (V3)--(V4);
          \draw (V4)--(V1);
          \draw (V1)--(V3);
        \end{tikzpicture}
      \end{center}
      \caption{\textsf{The distribution  of the branes, where the solid circles denote the ``$-$'' branes, the other two circles denote the ``+'' branes. Solid circles denote the branes where our real world resides.}} \label{Fig1}
      \end{figure}


\item[Subcase 2:] When there are only two branes $V_1=-V_2=3M^4$ in the RHS of (\ref{Brane-orig30}), we have
    \begin{equation}
    \tilde{\Lambda }=
    \frac{96M^4r^2_b}{(r^2-r^2_{b})^2}-24M^4{\left[\delta(y'-r'_{b})\delta(z'-r'_{b})
    -\delta(y'+r'_{b})\delta(z'-r'_{b})\right]}. \label{Subcase2}
    \end{equation}
    This exactly describes a degeneration to the usual $5$-dimensional RS model, except that the cosmological parameter is no more a constant; this confirms that our proposed theory is a generalization of the RS model. Moreover, it can be checked that in this framework the hierarchy problem is able to be alleviated as well.

\end{enumerate}

\end{enumerate}


\begin{section}
{Conclusion}
\end{section}

In this paper a six dimensional generalized RS model is studied, with a warp function $\sigma(y,z)$
depending on the two extra dimensions $y$ and $z$. No limits are artificially put on these two extra dimensional coordinates.

In contrast with the usual $5$-dimensional RS model, we derive a new condition, Eq.(\ref{ConformalFact-1}), which serves as a nonlinear equation. One solution is found; detailed discussion for it is as follows.

Firstly, when brane structures are absent, the distribution of the cosmological parameter on the plane of extra dimensions forms a family of concentric circles with radii expressed by the cosmological parameter,
$r = 8\sqrt { - \frac{M^4}{3\tilde {\Lambda }}}$. Here the circles are derived from the Einstein equation, instead of being added by hand as in the RS model.

Secondly, when 3-branes are present in the six-dimensional specetime, there are two cases:
\begin{enumerate}
\item[(i)] Singular points are absent. In this case, a new relationship between the cosmological parameter and the radius, Eq.(\ref{Case1}), is achieved.

\item[(ii)] Singular points are present. There are two subcases:

\begin{enumerate}

\item Generic case: It is found that the branes must reside at the singular points. By means of complex analysis, we expand the LHS of (\ref{Brane-orig29}) in the neighborhood of the singular points, and compare them to the RHS of (\ref{Brane-orig30}). Remarkably we discover that there must exist four branes symmetrically located on a circle, the tensions of which are $V_1=V_3=-V_2=-V_4=3M^4$.

\item Special case: There are only two branes, and (\ref{Subcase2}) yields a degeneration of our model to the usual 5D RS model.

\end{enumerate}

\end{enumerate}

Finally, it should be pointed out that the $S^1/\mathbb{Z}_2$ orbifold in the 5D RS model is a hyperthesis; but in our scenario it is able to be strictly derived. This demonstrates the advantage of our two-extra dimensional model. The methods and techniques developed in this paper may find important applications in relevant areas of Refs.\cite{Huangetal2013,ZhangHuang2011}\\

\section{Acknowledgment}

The authors are grateful to Profs. R.G. Cai and Y.-S. Piao for useful discussions. This work was supported by National Natural Science Foundation of China (No.11275017 \& 11173028).

\end{document}